# Explainable Text Classification in Legal Document Review

## A Case Study of Explainable Predictive Coding

Rishi Chhatwal, Esq.
Legal
AT&T Services, Inc.
Washington DC, USA
rishi.chhatwal@att.com

Peter Gronvall
Data & Technology
Ankura Consulting Group, LLC
Washington DC, USA
peter.gronvall@ankura.com

Nathaniel Huber-Fliflet
Data & Technology
Ankura Consulting Group, LLC
Washington DC, USA
nathaniel.huber-fliflet@ankura.com

Robert Keeling, Esq.
Complex Commercial Litigation
Sidley Austin LLP
Washington DC, USA
rkeeling@sidley.com

Dr. Jianping Zhang
Data & Technology
Ankura Consulting Group, LLC
Washington DC, USA
jianping.zhang@ankura.com

Dr. Haozhen Zhao
Data & Technology
Ankura Consulting Group, LLC
Washington DC, USA
haozhen.zhao@ankura.com

*Abstract—* In today's legal environment, lawsuits and regulatory investigations require companies to embark upon increasingly intensive data-focused engagements to identify, collect and analyze large quantities of data. When documents are staged for review – where they are typically assessed for relevancy or privilege – the process can require companies to dedicate an extraordinary level of resources, both with respect to human resources, but also with respect to the use of technology-based techniques to intelligently sift through data. Companies regularly spend millions of dollars producing 'responsive' electronically-stored documents for these types of matters. For several years, attorneys have been using a variety of tools to conduct this exercise, and most recently, they are accepting the use of machine learning techniques like text classification (referred to as predictive coding in the legal industry) to efficiently cull massive volumes of data to identify responsive documents for use in these matters. In recent years, a group of AI and Machine Learning researchers have been actively researching Explainable AI. In an explainable AI system, actions or decisions are human understandable. In typical legal 'document review' scenarios, a document can be identified as responsive, as long as one or more of the text snippets (small passages of text) in a document are deemed responsive. In these scenarios, if predictive coding can be used to locate these responsive snippets, then attorneys could easily evaluate the model's document classification decision. When deployed with defined and explainable results, predictive coding can drastically enhance the overall quality and speed of the document review process by reducing the time it takes to review documents. Moreover, explainable predictive coding provides lawyers with greater confidence in the results of that supervised learning task. The authors of this paper propose the concept of explainable predictive coding and simple explainable predictive coding methods to locate responsive snippets within responsive documents. We also report our preliminary experimental results using the data from an actual legal matter that entailed this type of document review. The purpose of this paper is to demonstrate the feasibility of explainable predictive coding in the context of professional services in the legal space.

## I. INTRODUCTION

In modern litigation, attorneys often face an overwhelming number of documents that must be reviewed and produced over the course of the matter. The costs involved in manually reviewing these documents has grown dramatically as more and more information is stored electronically. As a result, the document review process can require an extraordinary dedication of resources: companies routinely spend millions of dollars sifting through and producing responsive electronically stored documents in legal matters [1].

Attorneys have responded to the exponential growth of documents by employing machine learning techniques like text classification to efficiently cull massive volumes of data to identify responsive information. In the legal domain, text classification is typically referred to as predictive coding or technology assisted review (TAR). Predictive coding applies a supervised machine learning algorithm to build a predictive model that automatically classifies documents into predefined categories of interest. Attorneys typically employ predictive coding to identify so-called "responsive" documents, which are materials that fall within the scope of some 'compulsory process' request, such as discovery requests, subpoenas, or internal investigations.

Over the past few years, the authors of this paper have helped their clients reduce the cost of document review by millions of dollars across dozens of matters using predictive coding. These cost savings examples are not uncommon in legal matters, and in fact the cost-savings aspect to this technology has helped drive its adoption across the legal community. While predictive coding is regularly used to reduce the discovery costs of legal matters, it also faces a perception challenge: amongst lawyers, this technology is sometimes looked upon as a "black box." Put simply, many attorneys do not understand the underlying technology used to generate predictive models and to rank documents as responsive to a document request.

In 2016 this group of authors performed a study to address that challenge; we set out to share our understanding of the underlying technology with the legal community, through an illustrative research paper [2]. This research began to bridge the gap between academic research and the legal technology industry. Our view was that it is imperative that attorneys can reference peer reviewed research that thoroughly examines and measures the performance of the fundamental components of commonly used text analytics techniques. Without this knowledge, a legal team could find itself spending far too much time and money on a document review, to their client's detriment, by ineffectively implementing a predictive coding review protocol. A sound understanding of predictive coding's machine learning algorithms and preprocessing parameters places attorneys in a better position to control the costs and impact that it can have on a legal matter.

Legal technology experts continue to perform research to make predictive coding less of a "black box" in the minds of the legal community, but there is still more work to do [2]. The research in this study addresses a still misunderstood component of the predictive coding process: explaining how responsive documents are classified. Attorneys typically want to know why certain documents were determined to be responsive by the model, but sometimes those answers are not obvious or easily divined. In many instances, for example, a predictive model could be inaccurate, resulting in the model flagging irrelevant documents as highly likely to be responsive. This can be confusing for an attorney, especially if the text content of the document doesn't appear to contain obviously responsive content. An attorney with extensive knowledge of the training documents can make an educated guess as to why a document was highly-scored, but it can be difficult to pinpoint exactly what text in a document heavily influenced the high score. The focus of our research in this paper was to develop experiments that could target and examine the document text that the predictive model identified and used to generate the high score.

The Artificial Intelligence (AI) community has been researching explainable Artificial Intelligence since the 1970s, e.g. medical expert system MYCIN [3]. More recently, DARPA proposed a new direction for furthering research into Explainable AI (XAI) [4]. In XAI systems, actions or decisions are human understandable – "machines understand the context and environment in which they operate, and over time build underlying explanatory models that allow them to characterize real world phenomena." Similarly, in an explainable machine learning system, predictions or classifications generated from a predictive model are explainable and human understandable. Interpreting the decision of the machine is more important now than ever because, increasingly, machine learning systems are being used in human decision-making applications and machine learning algorithms are becoming more complex.

Understanding a model's classification decision is challenging in text classification because of the factors a model considers during the decision-making process, including word volume within text-based documents, and the volume and diversity of tokens established during the text classification process. In the legal domain, where documents can range from one-page emails to spreadsheets that are thousands of pages long, the complexity of the models creates challenges for attorneys to pinpoint where the classification decision was made within a document.

We find that the easiest way to understand a model's document classification decision is by evaluating small text snippets and identifying the ones that support the classification under examination. For the purposes of this paper, we considered a text snippet to be a small passage of words within a document usually ranging from 50 to 200 words. In legal document review, a document can be considered responsive when one or more of its text snippets are responsive and contain relevant information. Therefore, if predictive coding could locate these responsive snippets, attorneys could easily evaluate the model's document classification decision. In this scenario, it would be simple to explain why the model made its classification decision and this would help further minimize the black box nature of predictive coding. In addition to creating an explainable result, explainable predictive coding could enhance the overall document review process by reducing the time to review documents. Consider a scenario where attorneys only need to focus on the review of a responsive text snippet and not the entire text of a document – this would significantly speed up the review process and decrease the cost of the review.

Explainable predictive coding is well-suited for the legal document review process. It could also help improve investigative scenarios by quickly pinpointing potentially sensitive responsive content and enable a quick summary review of a small number of high-scoring documents snippets. Quickly understanding the content in a data set equips attorneys with the information they need to effectively represent their clients.

This paper proposes the concept of explainable predictive coding and simple explainable predictive modeling methods as an effective means to locate responsive snippets within a responsive document. We report our preliminary experimental results using the data from an actual legal matter, now concluded. The purpose of the paper is to demonstrate the feasibility of explainable predictive coding.

Section I introduces this concept. Section II discusses previous work in explainable text classification. We introduce our explainable predictive modeling method in Section III and report the experimental results in Section IV. We summarize our findings and conclude in Section V.

## II. PREVIOUS WORK IN EXPLAINABLE TEXT CLASSIFICATION

Research in Explainable AI has focused on two main areas of explainable machine learning: model-based explanations and prediction-based explanations. In a model-based approach, models are inherently explainable, and these types of models use decisions trees and If-Then rules to explain the results. Complex models, such as deep learning models like multilayer neural network models, non-linear SVM, or ensemble models are not directly human understandable and require implementing a more sophisticated approach to interpret the decision. With complex models, the proxy model approach is applied to create an explainable model that approximates the predictions of a given complex model [5].

An alternative to generating explainable models is to produce an explanation for each individual prediction generated by a complex model. Generally, a prediction-based explanation method provides an explanation as a vector with real-valued weights, each for an independent variable (feature), indicating the extent to which it contributes to the classification. This approach is not ideal for text classification, because of the high dimensionality in the feature space. In text classification, a document belongs to a category, mostly likely because some parts of the text in the document support the classification. Therefore, a small portion of the document text is often used as evidence to justify the classification result in text classification.

Predictive coding is an application of text classification to documents falling within the scope of a legal matter. Text data (documents) are often represented using the bag-of-words approach and characterized with tens of thousands of variables (words and/or phrases). Due to the high dimensionality, understanding the classifications made by text categorization models is very difficult and creates an interesting research opportunity. Recent research found that a prediction-based approach is often used to identify snippets of text as an explanation for the classification of a document. A text snippet that explains the classification of a document is called a 'rationale' for the document in [6].

Zaidan, et al. [6] proposed a machine learning method to use annotated rationales in documents to boost text classification performance. In their method, in addition to full document class labels, human document reviewers also highlighted the text snippets that explained why the corresponding document belonged to the class. The labeled documents together with their annotated text snippets were used as training data to build a text classification model using SVM. Their experiments showed that classification performance significantly improved with annotated rationales over the baseline SVM variants using an entire document's text.

Zhang, et al. [7] presented a Convolutional Neural Network (CNN) model for text classification that jointly employed labels on both documents and their constituent sentences. Specifically, they considered scenarios in which reviewers explicitly labeled sentences that support their overall document classification. Their method employed a two-level learning approach in which each document was represented by a linear combination of the vector representations of its component sentences. In the first level, a sentence-level convolutional model is built to estimate the probability that a given sentence supports a document-level classification. Then, in the second level, the CNN model leveraged the contribution of each sentence to the aggregate document representation in proportion to these probability estimates. Zhang, et al.'s approach was applied to five data sets that had document-level labels and the requisite sentence-level labels. Their CNN model experiments demonstrated that their approach consistently outperforms strong baselines. Moreover, their approach naturally provides explanations for the model's predictions because each sentence is evaluated for its probability of supporting the document-level classification.

In [8], Martens and Provost described a method in which the explanation of a document classification is a minimal set of the most relevant words, such that removing all the words in the set from the document would change the classification of the document. An algorithm for finding this minimal set of words was presented and they conducted case studies demonstrating the value of the method using two sets of document corpora.

Lei et al. [9] proposed an approach to generate rationales for text classifications. Their approach combined two components, a rationale generator and a rationale encoder, which were trained to operate together. The generator specified a distribution over text fragments (text snippets) as candidate rationales and the encoder decided the classifications of candidate rationales established by the generator. The proposed approach was evaluated on multi-aspect sentiment analysis against manually annotated test cases. The results showed that their approach outperformed the baseline by a significant margin. The approach was also successfully applied to a question retrieval task.

## III. EXPLAINABLE PREDICTIVE CODING

The main goal of the explainable predictive coding concept introduced in this paper is not to revisit underlying enhanced predictive model performance metrics, such as precision and recall. Rather, the goal is to provide additional information (explanations) for the labeling of each responsive document and to help attorneys more effectively and efficiently identify responsive documents during legal document review. As with many other explainable text classifications approaches, we use the prediction-based approach instead of the model-based approach. Also, we are interested only in generating explanations for responsive documents, therefore we focus on documents identified as responsive.

Explainable predictive coding sets out to build a model to estimate $Pr(r = Rationale \mid x, y = Responsive) \cdot Pr(y = Responsive \mid x)$, where $x$ is a document, $y$ is the model-labeled designation of the document ('responsive' or 'not responsive,' for instance), and $r$ is a text snippet from $x$. The task consists of two separate subtasks:

- The first subtask is a traditional text classification task where a text classification model is built to identify responsive documents.

- The second subtask is to generate a text classification model which will be used to identify responsive rationales in each responsive document. And it is the identification of the rationale that underlies the quest for the 'explainability' that this study seeks to develop.

Table 1 outlines the process for identifying rationales of responsive documents. In the following subsections, we describe two approaches in building models for identifying rationales of responsive documents in detail. The first approach uses the document model (trained using labeled documents) to identify rationales, while the second one builds a rationale model using a set of manually annotated rationales identified by attorneys.

Table 1: The Process for Identifying Rationales

| | |
|---|---|
| 1. | Train a Document Model |
| 2. | Train a Rationale Model |
| 3. | Use the Document Model to identify responsive documents |
| 4. | Break each responsive document into a set of overlapping text snippets with n words |
| 5. | Apply the Rationale Model to score all text snippets of responsive documents |
| 6. | For each responsive document, select the top n scored snippets as the identified rationale(s) |

*A. Document Model Method*

In this section, we introduce a simple rationale generation method. In this method, we used a set of training documents, each of which was labeled as either responsive or not responsive by attorneys, and also a set of documents to be classified. As mentioned above, the approach consists of two phases. In the first phase, as is typical in traditional predictive coding, a model is generated using the set of training documents and the model is deployed to identify a set of potentially responsive documents, using the model's predictions. In the second phase, the same model is applied to generate one or more rationales for each of the identified responsive documents.

To generate the rationales for a responsive document, we break the document into a set of small overlapping text snippets. We then apply the document model to classify these snippets, on the spectrum of highly likely to be responsive, down to not responsive. As such, the model assigns a probability score between 0 and 1 to each text snippet. The $n$ text snippets with the highest scores are selected as the rationales for the classifications under examination.

Determining the optimal size of the text snippets is a difficult task because the size of the rationales is unknown in advance of generating the rationale. To solve this problem, we deployed an iterative approach to break documents into a set of overlapping text snippets, as follows: we break a document into relatively large snippets – for snippets receiving large probability scores, we continue to break them into smaller sizes. This process continues until probability scores stop increasing.

*B. Rationale Model Method*

In predictive coding, training documents are manually reviewed by attorneys and are coded as responsive or not responsive. For this study, when a training document was coded as responsive, we asked the coding attorney to indicate why it was responsive, and to so indicate by highlighting the text snippets that supported his/her coding decision. These highlighted snippets – much like a training documents – became the manually annotated rationales.

Similar to the document model method, a traditional predictive coding model is trained in the first phase to identify responsive documents. In the second phase, a rationale model is trained to generate one or more rationales for each identified responsive document. In training a rationale model, we use annotated rationales of responsive documents as responsive training rationales and a set of randomly generated text snippets from non-responsive documents as not responsive training rationales. To generate rationales for a responsive document, we first break the document into a set of overlapping text snippets and then apply the rationale model to determine which of the text snippets are rationales.

IV. EXPERIMENTS

In this section, we report our results on a large data set from a real legal document review. We describe the data set and the design of the experiments in Sections IV.A and IV.B. Results are reported in Section IV.C.

*A. The Data Set*

The data set was collected from an actual legal matter, now concluded, that contains documents including email, Microsoft Office documents, PDFs, and other text-based documents. It consists of 688,294 documents manually coded by attorneys as responsive or not responsive. Among the 688,294 documents, 41,739 are responsive and the rest are not responsive. For each of the responsive documents, a rationale was annotated by a review attorney as the justification for coding the document as responsive. In practical terms, most rationales are continuous words, phrases, sentences or sections from the reviewed and labeled documents. A few rationales contain words that are comments from the attorney and do not occur in the documents. Some rationales may consist of more than one text snippet, which occur in different parts of the document.

Annotated rationales have a mean length of 52 words, with a standard deviation of 112.5 words. 97.5% of these rationales have fewer than 250 words. To reduce the effect of outliers, such as very long or very short rationales, we limit our rationales to those with 10 or more words but fewer than 250 words and those that can be precisely identified in the data set – resulting in 23,791 responsive documents with annotated rationales in our population. These 23,791 documents established our responsive population, covering 57% of all the responsive documents in the above 688,294 population. Proportionally, we randomly selected 365,742 documents from the not responsive documents within the 688,294 population to define the not responsive population in our experiments.

*B. Experiment Design*

The purpose of these experiments was to study the feasibility of automatically identifying rationales for responsive documents with and without annotated rationales. We conducted two sets of experiments. In both sets of experiments, we built two types of predictive models, a document model and a rationale model. A document model was trained using documents with responsive and not responsive labels, whereas a rationale model was trained using responsive and not responsive labeled text snippets. A responsive labeled snippet was an annotated rationale, while a not responsive labeled text snippet was a randomly selected text snippet from a not responsive document. Not responsive snippets could also be selected from responsive documents, but we did not fully explore that within the confines of this study. Rather, we adopted a random process to sample not responsive snippets from not responsive documents. To ensure not responsive snippets have similar parameters as responsive snippets, we enforced two constraints in sampling a not responsive snippet from a not responsive document: (1) the length should be a random number between 10 and 250, which is the same snippet length range as the responsive snippet population; (2) the starting position of the not responsive snippet should be a random position between zero and the attorney reviewed document's length, minus the snippet's length from the first constraint. One not responsive text snippet was selected from each not responsive document.

The first set of experiments evaluated the performance of both the document and rationale models in classifying annotated responsive rationales from not responsive snippets randomly selected from the not responsive documents. In these experiments, both document and rationale models were evaluated using a test set comprised of annotated rationales and randomly selected not responsive snippets. Precision and recall were used as performance metrics. The first set of experiments were performed to test the performance of the models on the text snippets alone and provide multiple ways to interrogate the results of the modeling methods.

The second set of experiments simulate a real legal application scenario and apply both document and rationale models to responsive labeled documents to identify rationales that "explain" the models' responsive decision. In these experiments, a responsive document is divided into a set of overlapping snippets. Then, the models are applied to these snippets to identify rationales. We encountered the following question: how should this study decide if a snippet should be treated as a rationale or not? Our answer to that question in basic predictive coding principles: one simple way is to identify the snippet with the largest score in a document and consider that as the rationale for the document. Recall (the percentage of identified rationales) was used as the performance metric. An annotated rationale is correctly identified if it is included in the text snippet with the largest score.

The machine learning algorithm used to generate the models was Logistic Regression. Our prior studies demonstrated that predictive models generated with Logistic Regression perform very well on legal matter documents [2, 10]. Other parameters used for modeling were bag of words with 1-gram and normalized frequency [2]. The results reported in the next section are averaged over a fivefold cross validation.

*C. Results of the Experiments*

Figure 1 details the precision and recall curves for the document and rationale models in <u>discriminating annotated responsive rationales from not responsive text snippets</u>. The curves are the average of fivefold cross validation results. Each of the five document models in the fivefold cross validation was trained using an 80% random subset of the 23,791 responsive documents and 365,742 not responsive documents, while each of the five rationale models were trained using an 80% random subset of the 23,791 annotated rationales and 365,742 not responsive text snippets. In each fold, both the document and rationale models were tested using a 20% random subset of the 23,791 annotated rationales and 365,742 not responsive text snippets, i.e. on average, 4,758 annotated rationales and 73,148 not responsive text snippets. The responsive document rate is 6.5%.

The second set of experiments evaluated the document and rationale models' <u>ability to identify rationales of responsive documents</u>. In these experiments, both the document and rationale models were the same models that were used in the first set of experiments. We use snippets of the responsive documents as the candidate rationales. Each responsive document was divided into a set of *n* (50, 100, and 200) word snippets with *n/2* words overlapping between neighboring text snippets. Table 2 details the statistics of the text snippets for each snippet setting. Both document and rationale models were deployed to assign probability scores to each snippet of a responsive document. *M* (1, 2, 3, 4, 5) top scoring snippets were selected as the identified rationales for each model. An identified snippet is a true rationale if it overlaps with the annotated rationale identified by the attorney reviewer.

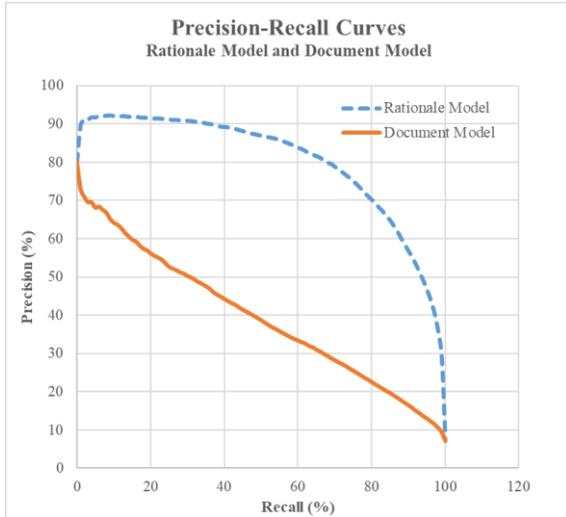

Figure 1: Precision and Recall Curves for Rationale Model and Document Model

Figure 1 shows that the rationale models performed quite well. At 80% recall, the rationale models achieved 70% precision. While performing less effectively than the rationale models, the document models' performance was encouraging, considering the responsive document rate is 6.5%. At 75% recall, the document models' achieved precision greater than 25%, which is roughly four times more than the responsive rate.

Table 2: Text Snippet Statistics

| Snippet Setting | Total Number of Snippets | Number of Documents | Average Number of Snippets |
|---|---|---|---|
| 50 | 933,997 | 23,791 | 39 |
| 100 | 473,181 | 23,791 | 20 |
| 200 | 242,578 | 23,791 | 10 |

Table 3 reports the recall (or percentage of responsive annotated rationales successfully identified) of the document and rationale models using different sizes of text snippets. The rationale models successfully identified responsive rationales for close to 50% of the responsive documents with the top scoring text snippet for all sizes of snippets. For the top three scoring snippets, the rationale models achieved more than 70% recall. For the top five scoring snippets, the rationale model achieved around 80% recall.

The rationale models performed better than the document models for snippets with 50 words. Given that the average rationale length in our experiment is 52 words (see Section IV.A), rationale models performed quite well in identifying snippets with a length similar to the rationales in the population. Both model types achieved similar recalls for snippets with 100 words, and the document models performed better than the rationale models for snippets with 200 words. One explanation for this observation is that a snippet rarely covers all words of the annotated rationale and a snippet almost always includes words that are not in the annotated rationale. Rationale models were trained using annotated rationales, therefore they did not tolerate noise (irrelevant words) very well. On the other hand, document models were trained using the full document text, which includes the words within the annotated rationales and also words throughout the rest of the document, thereby allowing it to be more tolerant of noise. Snippets with 200 words are typically longer than the average true rationale containing 52 words, thus may include a significant number of irrelevant words, which negatively impacts the performance of the rationale models. We observed that recall performance for the top scoring snippets degrades as the size of the snippets increases for the rationale models. This is likely because snippets with more words include more noise.

Table 3: Rationale Recall for Rationale and Document Models

| Number of words in Snippet | Top K Snippets | Rationale Recall | |
|---|---|---|---|
| | | Rationale Model | Document Model |
| 50 | 1 | **48%** | 44% |
| | 2 | **62%** | 56% |
| | 3 | **71%** | 65% |
| | 4 | **76%** | 71% |
| | 5 | **79%** | 75% |
| 100 | 1 | 47% | **51%** |
| | 2 | **64%** | 64% |
| | 3 | **73%** | 73% |
| | 4 | **79%** | 78% |
| | 5 | **82%** | 82% |
| 200 | 1 | 45% | **60%** |
| | 2 | 68% | **73%** |
| | 3 | 79% | **81%** |
| | 4 | 84% | **86%** |
| | 5 | 88% | **89%** |

While the results reported in this case study just begin to scratch the surface of how explainable predictive coding has practical applications in legal matters, they are promising: the results demonstrate that it is possible to build text classification models to identify rationales automatically, with or without the use of annotated rationales. In practical terms, this means that legal teams could evaluate responsive rationales generated by a Rationale Model or Document Model to substantially reduce the number of words an attorney must review to evaluate the responsiveness of a document.

**Rationale Model**

With this model and given that the average word count of a responsive document in this study is 970 words, an attorney who is reviewing the top four 50-word snippets (125 to 250 words, accounting for the overlap among snippets) could exclude, on average, 720 to 845 words from the review of each document, while still achieving 76% recall. Extrapolating these results across an entire set of responsive documents, the average word savings in these scenarios is 17,129,520 to 20,103,395 words per experiment. Using the average number of words per document to estimate the document review savings, the use of the Rationale Model could result in a savings of 17,659 to 20,725 documents — or 74% to 85% of the responsive documents.

**Document Model**

The top five Document Model snippets achieve a similar recall (75%) to that of the Rationale Model's top four 50-word snippets. In practice, a Document Model is easier to implement when compared to a Rationale Model because attorneys do not have to annotate rationales to create responsive training data. Using the Document Model, an attorney reviewing the first top 50-word snippet could exclude, on average, 920 words from the review of each document while still achieving 44% recall. These results are particularly compelling because an attorney would only have to evaluate nearly half of the responsive information in the data set and also exclude a substantial amount of irrelevant information. This approach would result an average savings of 21,887,720 words when extrapolated across the entire responsive document set for these experiments.

V. SUMMARY

Explainable AI has drawn the attention of the AI and legal communities alike. It has proven that it is possible to develop machine learning systems that are explainable, to meet practical demands and to effect positive impacts on legal outcomes. The authors believe that explainable AI has the potential to significantly advance the application of text classification in legal document review matters. This study fortified our thinking that the concept of explainable predictive coding has a promising future in the legal document review realm. These results open the door to conduct future studies in explainable predictive coding. We plan to conduct more experiments on additional data sets. Further, we plan to explore more advanced machine learning technologies to continue evolving our understanding of categorization rationales.

ACKNOWLEDGMENT

We thank Katie Jensen for helping us prepare the data set and we thank the anonymous reviewers for their helpful comments.